\documentclass[12pt]{iopart}
\usepackage{iopams,bm}
\usepackage{mathrsfs}
\usepackage{graphicx}

\newcommand{\FDR}{\ensuremath \Xi}
\newcommand{\FDRq}{\ensuremath \widehat\FDR}
\newcommand{\reff}[1]{(\ref{#1})}
\newcommand{\I}{{\it i}}
\newcommand{\II}{{\it ii}}
\newcommand{\III}{{\it iii}}
\newcommand{\IV}{{\it iv}}
\newcommand{\A}{A}
\newcommand{\B}{B}
\newcommand{\C}{C}
\newcommand{\D}{D}
\newcommand{\p}{\ensuremath \varphi}
\newcommand{\tp}{\ensuremath \tilde\varphi}
\newcommand{\OO}{\mathscr O}
\newcommand{\ha}{\ensuremath \hat a}

\newcommand{\vs}{\ensuremath \varsigma}
\newcommand{\FF}{\ensuremath {\mathcal F}}
\newcommand{\RR}{\ensuremath {\mathcal R}}
\newcommand{\AAA}{\ensuremath {\mathcal A}}
\newcommand{\G}{\ensuremath (0)}
\newcommand{\bt}{\ensuremath \bar \theta}
\newcommand{\eps}{\ensuremath\epsilon}
\newcommand{\f}{\mathfrak f}
\newcommand{\lsi}{{\rm LSI}}
\newcommand{\ft}{{\rm FT}}

\newcommand{\be}{\begin{equation}}
\newcommand{\ee}{\end{equation}}

\newcommand{\deff}{\ensuremath \mathrel{\mathop:}=}

\begin{document}

\title[Ageing in the contact process]{Ageing in the contact process:
Scaling behavior and universal features}

\author{Florian Baumann$^{1,2}$ and Andrea Gambassi$^{3,4}$}
\address{$^1$Institut f\"ur Theoretische Physik I,
Universit\"at Erlangen-N\"urnberg, \\
Staudtstra{\ss}e 7B3, D -- 91058 Erlangen, Germany}
\address{$^2$Laboratoire de Physique des
Mat\'eriaux,\footnote{Laboratoire associ\'e au CNRS UMR 7556}
Universit\'e Henri Poincar\'e Nancy I, \\
B.P. 239, F -- 54506 Vand{\oe}uvre l\`es Nancy Cedex, France}
\address{$^3$ Max-Planck-Institut f\"ur Metallforschung, \\
Heisenbergstra{\ss}e 3, D--70569 Stuttgart, Germany}
\address{$^4$ Institut f\"ur Theoretische und Angewandte
Physik, Universit\"at Stuttgart, \\ Pfaffenwaldring 57, 
D-70569 Stuttgart, Germany}

\begin{abstract}
We investigate some aspects of the ageing behavior observed in the
contact process after a quench from its active phase to the critical
point.
In particular
we discuss the scaling properties of the two-time response function
and we calculate it and its {\it universal ratio} to the
two-time correlation function 
up to first order in the
field-theoretical $\eps$-expansion ($\eps = 4 - d$). 
The scaling form of the response function does not fit the prediction
of the theory of local scale invariance.
Our findings 
are in good qualitative agreement with recent numerical
results.
\end{abstract}

\maketitle

\setcounter{footnote}{0}

\section{Introduction}

The contact process (CP) has been introduced long ago as a simple model for
the spreading of diseases~\cite{Harris-74} (see
also~\cite{janssen,tauber,hinrichsen1,Odor-04}). It is defined on a
$d$-dimensional hypercubic lattice in which each site $x$ is characterized
by an occupation variable $n_x$ and is either
active (ill, $n_x=1$) or inactive (healthy, $n_x=0$). 
At each time step of the dynamics a site $x$ is chosen at random. 
If $x$ is inactive then it becomes active with a rate given by
$\Lambda$ times the fraction of its neighbouring sites which are
already active (infection). Otherwise, 
if originally active, $x$ becomes inactive
with rate $1$ (healing). In the long-time (non-equilibrium) stationary state
all lattice sites will be eventually inactive ({\it absorbing phase})
if $\Lambda$ is small enough, whereas for
large enough $\Lambda$ a finite fraction of sites will
be active ({\it active phase}).
These two phases are separated by a continuous phase transition for
$\Lambda=\Lambda_c$ (with $\Lambda_c \simeq 3.3$ in $d=1$, see,
e.g.,~\cite{hinrichsen1}) and a suitable order parameter is indeed
provided by the average density of active sites
$\langle n_x \rangle$ in the stationary state. 
 
Extensive Monte Carlo simulations of the CP have provided a very
detailed quantitative description of its dynamical behavior and of the associated
non-equilibrium phase transition, 
which turns out to be in the same universality class as directed
percolation and therefore its universal features are properly captured
by the so-called {\it Reggeon field
theory} (see, e.g., \cite{janssen,tauber,janssen2,cardy}).

The directed percolation (DP) 
universality class has been recently the subject of renewed interest
in the context of ageing
phenomena~\cite{enss,ramasco,hinrichsen2,henkel06}, which are
characterized by two-time quantities (such as response
and correlation functions, see below) which are homogeneous functions
of their time arguments 
and do not display the time-translational invariance which
is expected in a stationary state.
The two-time quantities of interest
are the connected density-density correlation 
function $C_{x-x'}(t,s) \deff \langle n_x(t) n_{x'}(s) \rangle - \langle
n_x(t)\rangle\langle n_{x'}(s)\rangle$ and the response function
$R_{x-x'}(t,s)\deff \delta\langle n_x(t)\rangle/\delta
\kappa_{x'}(s)|_{\kappa=0}$ where $\kappa_{x'}(s)$ is the field 
conjugate to
$n_{x'}(s)$ --- corresponding to a local spontaneous activation rate of the 
lattice site $x'$ at time
$s$ --- and $\langle\ldots\rangle$ stands for the
average over the stochastic realization of the process. 

In~\cite{enss,ramasco} the ageing behavior of the 
critical CP was investigated
numerically. A system with a homogenous initial particle
density (corresponding to the active phase with $\Lambda \gg
\Lambda_c$) was quenched at $t=0$ 
to the critical point $\Lambda = \Lambda_c$
and $R_{x=0}(t,s)$ and $C_{x=0}(t,s)$
were determined, finding
the following scaling behavior ($t>s$)
\begin{equation}
\label{scaling_forms}
R_{x=0}(t,s) = s^{-1-a} f_R(t/s), \qquad C_{x =
0}(t,s)= s^{-b} f_C(t/s) ,
\end{equation}
where
\begin{equation}
\label{scaling_forms2}
f_R(y\gg1)\sim y^{-\lambda_R/z}, \qquad f_C(y\gg1)\sim
y^{-\lambda_C/z} 
\end{equation}
and $b=2\beta/(\nu z) = 2\delta$
[$\beta$, $\nu$, $z$ and  $\delta \deff \beta/(\nu
z)$ are the standard critical exponents of DP, see, e.g.,
\cite{hinrichsen1} --- the values of $z$
and $\delta$ for $d=1,2$ and $3$ are reported in
table~\ref{tabexp}]. In particular it was found that
$\lambda_C/z = 1.85(10)$ and
$\lambda_R/z = 1.85(10)$ for $d=1$~\cite{enss,ramasco}, whereas
$\lambda_C/z = 2.75(10)$ and $\lambda_R/z = 2.8(3)$ for
$d=2$~\cite{ramasco}. These results  suggest that $\lambda_R = \lambda_C$
independently of the dimensionality $d=1$ and $2$. On the same numerical footing it was
noticed --- with some surprise --- that $1+a = b$,
in stark contrast to the
case of slow dynamics of systems with detailed balance
such as Ising ferromagnets, for which $a=b$ (see, e.g.,~\cite{gl-00,cg-rev}). 
The fluctuation-dissipation ratio~\cite{fdr-93-94,gl-00}
\be
X(t,s)\deff T_b R_{x=0}(t,s)/\partial_s
C_{x=0}(t,s)
\label{defXDB}
\ee
and its long-time limit $X_\infty \deff
\lim_{s\rightarrow\infty}\lim_{t\rightarrow\infty} X(t,s)$ have been
used, for these magnetic systems evolving in contact
with a thermal bath of temperature $T_b$, to detect whether the
(equilibrium) stationary state has been reached, in which case the
fluctuation-dissipation theorem yields $X_\infty = 1$.
Given that $b = a +1$ for the CP, the fluctuation-dissipation ratio
--- as defined in equation~\reff{defXDB} --- would always yield a trivial value of $X_\infty$ and
therefore it would not serve its purpose.
In~\cite{enss}, it was suggested to consider, instead of $X$, 
\begin{equation}
\label{fdr}
\FDR(t,s) \deff \frac{R_{x=0}(t,s)}{
C_{x=0}(t,s)} =
\frac{f_R(t/s)}{f_C(t/s)}
\end{equation}
and define $\FDR_\infty \deff \lim_{s \rightarrow \infty}\lim_{t \rightarrow
\infty} \FDR(t,s)$, which has now a finite non-trivial value
$\FDR_\infty = 1.15(5)$ for the CP in $d=1$~\cite{enss}. (We
comment on the meaning of $\FDR$ in subsection~\ref{sec-relX}.)
In addition, the numerical results of~\cite{enss,ramasco} support the
applicability of the so-called theory
of local scale invariance (LSI)~\cite{henkel02,henkel04,picone}
to the ageing behavior of the CP.
This theory tries to use {\it local} space-time symmetries to
constrain the form of $f_R(t/s)$ and has been 
applied to ageing phenomena in magnetic systems quenched from an
initial high-temperature state to and below the critical
temperature~\cite{henkel04,picone} and bosonic
reaction-diffusion systems~\cite{baumann1}. Recent simulations suggest
that LSI holds also in the parity conserving universality class~\cite{odor}. 
However, in the case of critical ageing of the Ising model with purely
relaxational dynamics, it was shown 
that corrections to the prediction of LSI are present both in 
field-theoretical results at two loops~\cite{cg-rev,calabrese3} 
and in simulations~\cite{pleimling,lcz-06}.
A more general version of LSI~\cite{henkel06} --- the
version we shall refer to in this paper --- improved
considerably the agreement with simulations while the
disagreement with field-theoretical predictions remained.
Also in  the case of the CP, numerical results indicate that 
the scaling function $f_R$ predicted by LSI is
incorrect for $t\simeq s$~\cite{hinrichsen2,henkel06}. 
It was suggested in~\cite{henkel06,henkel06_rev} that
one could possibly account for this discrepancy by extending LSI 
to include also the case of a nonvanishing average of the
order parameter (in the present version a vanishing order
parameter is implicitly assumed). 
As this extension is at present still
lacking, in what follows we shall refer to the available version of LSI.

The previously mentioned  (numerical) works leave some questions open: 
\vspace{-0.7cm}
\begin{itemize}
\item[(\I)] Could the relations $1+a=b$ and $\lambda_R=\lambda_C$ have
been expected?
\item[(\II)] Is $\lambda_R $ ($= \lambda_C$) an independent critical
exponent? 
\item[(\III)] Is $\FDR_\infty$ a {\it universal} quantity like $X_\infty$
in ageing systems with detailed balance?
\item[(\IV)] To what extent does LSI describe some of the features of
the ageing behavior in the CP?
\end{itemize}
\vspace{-0.7cm}
The aim of this note is to answer these questions by adopting a
field-theoretical approach.

The rest of the paper is organized as follows.
In section~\ref{sec-FTA} we introduce the field-theoretical 
model (Reggeon field theory), the formalism
and we set up the general framework for the perturbative
expansion. In addition we discuss the expected scaling forms for the
two-time response and correlation functions, the
relation among the different ageing exponents and between $X$ and
$\FDR$, providing complete answers to the questions (\I), (\II) and
(\III).
In section~\ref{sec-RF}, we calculate the response function and
its long-time ratio $\FDR_\infty$ to the correlation function 
up to first order in the
$\epsilon$-expansion ($\epsilon=4-d$) and then we compare our
results to the predictions of LSI, providing an answer to question (\IV).
In section~\ref{sec-concl} we summarize our findings and present our
conclusions.
Some details of the calculation are reported in the appendix.

%
\section{The field-theoretical approach}
\label{sec-FTA}

As explained in detail in~\cite{janssen,tauber}, the universal scaling
properties of the CP (more generally, of the DP 
universality class) in the stationary state 
are captured by a Reggeon field-theoretical action $S$.
In the critical case it reads
\begin{equation}
S[\p,\tp] = \int\!\!\rmd^dx\,\rmd t\, \left\{ \tp \left[\partial_t -
D \nabla^2\right] \p - u (\tp-\p)\tp \p  - h \, \tp\right\}\,,
\label{action}
\end{equation}
where the $\p(x,t)$ and $h(x,t)$ are the coarse-grained versions of 
the particle density $n_x(t)$ and of the 
spontaneous activation rate of lattice sites 
$\kappa_x(t)$ (external perturbation), respectively, whereas
$\tp(x,t)$ is the response field, $u>0$ the bare coupling constant of
the theory and $D$ the diffusion coefficient ($D=1$ in what follows,
unless differently stated). 

In terms of the action~\reff{action}, 
the average of an observable $\OO$ over the possible stochastic
realizations of the process is given by $\langle\OO\rangle =
\int [\rmd \p\rmd\tp]\, \OO \, \rme^{-S[\p,\tp]}$. As a result, 
the response of $\langle\OO\rangle$
to the external perturbation $h$ can be computed as
$\delta\langle\OO\rangle/\delta h(x,t) =
\langle\tp(x,t)\OO\rangle$, leading to the following expression for
the (linear) response function $R_{x-x'}(t,s) \deff \delta
\langle \p(x,t)\rangle/\delta h(x',s)|_{h=0} = \langle
\p(x,t)\tp(x',s)\rangle_{h=0}$.

The action~\reff{action} with $h=0$ and $t\in (-\infty,\infty)$ 
is invariant under the duality
transformation
\be
\tp(x,t) \stackrel{{\rm RR}}{\longleftrightarrow} -\p(x,-t)
\label{eq-RR}
\ee
(the so-called {\it rapidity reversal} --- RR) 
which implies that the scaling dimensions $[\ldots]_{\rm scal}$ 
of the fields $\p$ and $\tp$ are equal (see, e.g, \cite{janssen,tauber}): 
\be
[\p(x,t)]_{\rm scal} = [\tp(x,t)]_{\rm
scal} = \beta/\nu \;. 
\ee 
Note that, as in the case of systems with detailed balance (DB), the scaling
dimensions of the fields (with $t>0$) do not change if RR (alternatively, DB)
is broken by the presence of {\it initial conditions} (at time $t=0$).\\
If RR is a symmetry of $S$, then 
\begin{eqnarray}
\hspace{-1cm}
C_{x-x'}(t,s) &\deff& \langle
\p(x,t)\p(x',s)\rangle_{h=0} - \langle\p(x,t) \rangle_{h=0}
\langle\p(x',s) \rangle_{h=0} \nonumber \\
&\stackrel{{\rm RR}}{=} &\langle
\tp(x,-t)\tp(x',-s)\rangle_{h=0}  - \langle\tp(x,-t) \rangle_{h=0} \langle
\tp(x',-s)\rangle_{h=0} = 0 \, ,
\label{cc0}
\end{eqnarray} 
where the last equality is a
consequence of causality~\cite{zinn-justin}, and therefore correlations
vanish in the stationary state of the CP. 
On the other hand, a suitable initial condition (say, at time $t=0$) 
can effectively generate correlations which decay for $t>0$.
Generally speaking one expects this decay to be exponential in
$t$ for $\Lambda\neq\Lambda_c$, due to a finite relaxation time both
in the absorbing and active phase, whereas an algebraic decay is
expected in the critical case $\Lambda=\Lambda_c$ we are interested in.
(This picture is indeed confirmed by the numerical results of~\cite{enss}.)
Accordingly, $C_{x=0}(t,s) \deff \langle \p(x,t)\p(x,s)\rangle - \langle
\p(x,t)\rangle \langle\p(x,s)\rangle $ does no
longer vanish during the relaxation from the initial condition 
and for $t$, $s\neq 0$ it has the same scaling dimension as
$R_{x=0}(t,s)$. 
Two straightforward consequences of this fact are:
\vspace{-0.7cm}
\begin{itemize}
\item[(\A)] $1+a=b$ [see equation~\reff{scaling_forms}], 
which could have been expected on this basis and therefore is
valid beyond the mere numerical coincidence\footnote{The same conclusion can
be drawn from the analysis presented in~\cite{oerding} (see
also~\cite{janssen,tauber})} and 
\item[(\B)] $\FDR(t,s)$ [see
equation~\reff{fdr}] --- and therefore $\FDR_\infty$ --- 
is a ratio of two quantities with the same
engineering and scaling dimensions and therefore it 
is a {\it universal} function (see,
e.g.,~\cite{pha-91}) which takes the same value in all the models
belonging to the same universality class (in particular, the lattice
CP, the DP and the model~\reff{action} on the
continuum). This is analogous to the case of $X_\infty$ in
systems with DB.
\end{itemize}

\subsection{Scaling forms}
\label{sec-scalingforms}

The non-equilibrium dynamics of the CP after a sudden quench (at time
$t=0$) from the active phase $\Lambda \gg \Lambda_c$ (characterized by
$\langle n_x(t<0) \rangle = 1$ on the lattice and $\langle
\p(x,t<0)\rangle \neq 0$ on the continuum) to the critical point
$\Lambda = \Lambda_c$ is partly analogous to 
the non-equilibrium relaxation of Ising
systems with dissipative dynamics after a quench from a magnetized
state (e.g., a low-temperature one with $T\ll T_c$, where $T_c$ is the
critical temperature) to the critical point
$T=T_c$~\cite{calabrese1}. In both cases the order parameter
$m(t)\deff \langle \p(x,t) \rangle$\footnote{For magnetic systems $\p$
represents the local magnetization whereas in the present case 
$\langle \p \rangle$ is the coarse-grained density of active sites.} 
\footnote{Hereafter we assume invariance of the system under space
translations.} provides a background for the fluctuations,
with a {\it universal} scaling behavior~\cite{vWOH-98,jss}
\be
m(t) = A_m m_0 t^{\theta + \ha} {\mathcal F}_M(B_m m_0
t^\vs)
\label{scalingm}
\ee
where $m_0\deff \langle \p(x,t=0)\rangle$ 
is the initial value of the order parameter, 
$\ha$ is
related to the scaling dimensions of the fields $\p$ and $\tp$ in real space via
$\ha = (d - [\p]_{\rm scal} - [\tp]_{\rm scal})/z$ (we recall that
$[\p]_{\rm scal} = \beta/\nu$), $\theta$ is the
so-called initial-slip exponent~\cite{jss} 
and $\vs = \theta + \ha + \beta/(\nu z)$. In
equation~\reff{scalingm}, $\FF_M(v)$ is a universal scaling function 
once the non-universal amplitudes $A_m$ and $B_m$ have been fixed by
suitable normalization conditions, e.g., $\FF_M(0)=1$ and
$\FF_M(v\rightarrow \infty) = v^{-1} + O(v^{-2})$. [Note that
$\FF_M(v\rightarrow\infty)\sim v^{-1}$ is required in order 
to recover the
well-know long-time decay of the order parameter $m(t)\sim
t^{-\beta/(\nu z)}$.] Within the Ising
universality class $\ha = (2-\eta-z)/z$ and $\theta \neq
0$~\cite{jss,cg-rev} whereas within the DP 
class $\ha = -\eta/z$ (indeed $[\p]_{\rm
scal} = [\tp]_{\rm scal} = (d+\eta)/2$~\cite{janssen}) and $\theta =
0$~\cite{vWOH-98}.
In both cases it turns out that the width $\Delta_0 = \langle [\p(x,0)
- m_0 ]^2\rangle$ of the initial distribution 
of the order parameter $\p$ --- assumed with short-ranged spatial
correlations --- is irrelevant~\cite{jss,vWOH-98} and controls
only corrections to the leading scaling behavior, so that
an initial state with $\Delta_0 \neq 0$ is
asymptotically equivalent
to one with no fluctuations $\Delta_0 = 0$.
Equation~\reff{scalingm} clearly shows that a non-vanishing value of
the initial order parameter introduces an additional time scale in the
problem $\tau_m = (B_m
m_0)^{-1/\vs}$~\cite{janssen,vWOH-98,calabrese1}. The long-time limit
we are interested in is characterized by ${\rm times} \gg \tau_m$ and
therefore the relevant scaling properties can be formally explored in
the limit $m_0 \rightarrow \infty$, i.e., $\tau_m \rightarrow 0$.

According to the analogy discussed above one expects the 
following scaling form for the two-time response function in
momentum space (see, e.g., \cite{janssen,vWOH-98} and section 3
in~\cite{calabrese1})
\be
R_{q=0}(t>s,s) = A_R\, (t-s)^{\ha}(t/s)^\theta F_R(s/t,B_m m_0^{1/\vs} t)\; ,
\label{scalR}
\ee
where $F_R$ is a universal scaling function once the non-universal
amplitude $A_R$ has been fixed by requiring, e.g., $F_R(0,0) = 1$.
In the case of the DP universality class, the two-time
Gaussian correlation function $C^{\G}_x(t,s)$ vanishes for $m_0 = 0$ --- apart from
irrelevant corrections due to a finite $\Delta_0$ --- whereas the
two-time Gaussian response function $R^{\G}_x(t,s)$ is invariant under
time translations. Taking into account causality, it is easy to realize
that all the diagrammatic contributions to the full response function
(with $u\neq 0$) have the same invariance and therefore $F_R(s/t,0) =
1$. (As opposed to the case of the Ising universality class, where
$\theta\neq 0$ and 
$F_R(s/t,0)\neq 1$~\cite{cg-02a,calabrese3,cg-rev,pleimling}.) In fact,
the case $m_0=0$ would correspond on the lattice to an initially empty
system with $\Lambda = \Lambda_c$ where nothing happens as long as
one does not switch on the external field $\kappa_{x'}(s)$. Accordingly, the
response to  $\kappa_{x'}(s)$, at time $s + \Delta t$ has to depend
only on $\Delta t$. 

In the long-time limit $t>s \gg \tau_m$ one expects $F_R$ to become
independent of the actual value of $m_0$ and therefore
\be
F_R(x,v\rightarrow\infty) = \frac{\AAA_R}{A_R} x^{\theta - \bt} \FF_R(x) ,
\label{frasy}
\ee
where $\bt$ is an additional exponent [related to $\lambda_R$, cf
equation~\reff{lambdas}], so that the resulting scaling is
\be
R_{q=0}(t>s,s) = \AAA_R\, (t-s)^{\ha}(t/s)^{\bt} \FF_R(s/t)\; .
\label{scalRasy}
\ee
$\AAA_R$ is a {\it non-universal} constant which can be fixed by
requiring $\FF_R(0)=1$ and which has a {\it universal} ratio to $A_R$
[see equation~\reff{frasy}]. 
For the Ising universality class, is was found~\cite{calabrese1}
\be
\bt = - \left( 1 + \ha + \frac{\beta}{\nu z} \right) .
\label{expbt}
\ee
On the other hand, it is easy to see that the scaling arguments which
were used to drawn this conclusion for the response function
(see section 3 in~\cite{calabrese1}) are valid also for the
CP. 

In~\cite{oerding} the scaling behavior of the 
two-time correlation function $C_{q=0}(t,s)$, after a quench from the
active phase to the critical point, has been discussed with the result
that (see equations~(3.3) and (4.14) in~\cite{oerding})
\be
C_{q=0}(t>s,s) = \AAA_C (t-s)^{\ha} (t/s)^{\bt} \FF_C(s/t)
\label{scalCasy}
\ee
where the non-universal amplitude $\AAA_C$ can be fixed by requiring
$\FF_C(0) = 1$, and $\bt$ is indeed given by equation~\reff{expbt}. 
Note that in analogy with $\FDR$ [see equation~\reff{fdr}] one can
also define (we assume $t>s\gg\tau_m$)
\be
\FDRq(t,s) \deff \frac{R_{q=0}(t,s)}{C_{q=0}(t,s)} =
\frac{\AAA_R}{\AAA_C}\frac{\FF_R(s/t)}{\FF_C(s/t)}
\label{fdrq}
\ee 
[where we have used the scaling forms~\reff{scalRasy} and~\reff{scalCasy}] and
\be
\FDRq_\infty \deff \lim_{s\rightarrow\infty }\lim_{t\rightarrow\infty}
\FDRq(t,s) = \frac{\AAA_R}{\AAA_C}
\label{fdrqinf}
\ee
which are, as $\FDR(t,s)$ and $\FDR_\infty$, a {\it universal}
function and amplitude ratio, respectively. Although in general
$\FDR(t,s) \neq \FDRq(t,s)$, the argument discussed
in~\cite{cg-02a,cg-rev} can be used to conclude that $\FDR_\infty =
\FDRq_\infty$ also in this case. \hfill\\

In what follows we focus on $R_{q=0}$ and $C_{q=0}$, i.e., the
response and correlation function of the spatial average of the
density of active sites ($N^{-1} \langle \sum_x n_x \rangle$ on a
lattice with $N$ sites). On the other hand, the corresponding scaling
forms equation~\reff{scalRasy} and~\reff{scalCasy} can be easily
generalized to $q\neq 0$ taking into account that this
amounts to the
introduction of an additional scaling variable $y = A_D D q^z (t-s)$
where $A_D$ is a dimensional non-universal constant which can be fixed
via a suitable normalization condition. Accordingly, the
scaling behavior of the autoresponse $R_{x=0}$ and autocorrelation
$C_{x=0}$ functions can be easily worked out from equations~\reff{scalRasy}
and~\reff{scalCasy}, 
leading to the
identification of the exponents $a$ and $\lambda_{R,C}$ in
equations~\reff{scaling_forms} and~\reff{scaling_forms2} as
\be
1 + a = b = \frac{d}{z} - \ha  = \frac{2 \beta}{\nu z} = 2\delta
\ee
and
\be
\frac{\lambda_C}{z} = \frac{\lambda_R}{z} = \frac{d}{z} - \ha - \bt =
1 + \delta + \frac{d}{z} .
\label{lambdas}
\ee
Therefore we conclude that:
\vspace{-0.7cm} 
\begin{itemize}
\item[(\C)] $\lambda_R$ is {\it equal} to $\lambda_C$ beyond the mere numerical
coincidence observed in~\cite{enss,ramasco} and
\item[(\D)] $\lambda_R = \lambda_C$ is {\it not} an independent
critical exponent but it is given by 
\be 
\lambda_R = z + z \delta + d\,.
\label{eq-exp}
\ee 
In table~\ref{tabexp} we compare the values of $\lambda_{R,C}$
obtained from this scaling relation and the available 
estimates of $\delta$ and $z$ to the results of fitting 
the asymptotic behavior of the scaling functions $f_{C,R}$ [see
equations~\reff{scaling_forms} and~\reff{scaling_forms2}] from
numerical data~\cite{enss,ramasco}. The results
reported in the last three columns are in quite good agreement 
both for $d=1$ and $2$.
\end{itemize}
\vspace{-0.7cm}

\begin{table}[h]
\begin{center}
\begin{tabular}{|c|l|l|l|l|l|} 
\hline
$d$ & $\delta$ & $z$ & $1 + \delta + d/z$ & $\lambda_R/z$ & $\lambda_C/z$ \\
\hline
$1$ & $0.159464(6)$ & $1.580745(10)$  & $1.792077(10)$ & \begin{minipage}{2cm}
$1.85(10)$~\cite{enss} \\ $1.9(1)\phantom{22}$~\cite{ramasco} \\
$1.76(5)$~\cite{hinrichsen2} \end{minipage} & \begin{minipage}{2cm}
$1.85(10)$~\cite{enss} \\ $1.9(1)\phantom{22}$~\cite{ramasco}\\ \phantom{$1.76(5)$~\cite{hinrichsen2}}\end{minipage}\\%
\hline
$2$ & $0.451$ & $1.76(3)$ & $2.58(2)$ & $2.75(10)$~\cite{ramasco} &
$2.8(3)\phantom{22}$~\cite{ramasco} \\%
\hline
$3$ & $0.73$ & $1.90(1)$ & $3.30(1)$ &\multicolumn{2}{c|}{not available} \\
\hline
\end{tabular}
\caption{%
Comparison between the direct numerical estimates of
$\lambda_{R,C}/z$ and the predictions of the scaling relation
$\lambda_R/z (=\lambda_C/z) = 1 + \delta + d/z$ [see
equation~\reff{lambdas}] in which the available estimates of 
$\delta$ and $z$ (taken from
table~2 in~\cite{hinrichsen1}) are used. 
The values of $\lambda_{R,C}/z$
reported in the last two columns have been obtained by fitting the scaling
behavior of the autoresponse and autocorrelation function determined
via density-matrix
renormalization-group computation~\cite{enss} and by  Monte Carlo
simulations~\cite{ramasco,hinrichsen2} of the contact process. 
\label{tabexp}
}
\end{center}
\end{table}

The conclusions (\A--\D) we have drawn so far provide complete
answers to the questions (\I--\III) which we have posed at the end of the
Introduction. 

\subsection{Relation between $X$ and $\FDR$}
\label{sec-relX}
 
In~\cite{enss} $\FDR$ has been introduced on the basis of a formal analogy
with the fluctuation-dissipation ratio $X(t,s)$. 
Here we argue that indeed $X$ and $\FDR$ play the same role in
different circumstances.
 
The stationary state of a system with detailed balance is
characterized by the time-reversal (TR) symmetry of the dynamics,
which qualifies the state as an {\it equilibrium} one with a
certain temperature $T$. Time-translational invariance (TT) and TR
symmetry of the dynamics in the 
equilibrium state imply the fluctuation-dissipation
theorem $T R_{x,q}(t,s) = \partial_s C_{x,q}(t,s)$, leading to $X =
1$ [see equation~\reff{defXDB}]. In this sense $X\neq 1$ 
is a signature of the fact that the system has not reached its stationary
(equilibrium) state. (This is usually due to slowly relaxing modes
which prevent the system from ``forgetting'' the initial conditions of
the dynamics~\cite{cu-rev,cg-rev}.)  
When detailed balance does not hold, TR is no longer a symmetry of the
stationary state. %
In the specific case of the DP universality class, however, 
the {\it non-equilibrium}
stationary state is characterized by a different symmetry, the
rapidity-reversal [RR, see equation~\reff{eq-RR}], 
which leads --- as discussed in
subsection~\ref{sec-scalingforms}, equation~\reff{cc0} 
--- to vanishing correlations and therefore
to $\FDR^{-1},\FDRq^{-1} = 0$. 
Accordingly, $\FDR^{-1},\FDRq^{-1} \neq 0$ provides
a signature of the fact that the system is not in its
stationary state. In this sense $\FDR$ ($\FDRq$) is analogous to $X$ for
both of them indicate if the stationary state is eventually reached. 
In particular this occurs {\it generically} after a perturbation
(e.g., a sudden change in the temperature $T$ or the spreading rate $\Lambda$)
and therefore $X_\infty = 1$ or $\FDR_\infty^{-1} = \FDRq_\infty^{-1}
= 0$. On the other hand, it may happen that for particular choices
of the external parameters (e.g., $T$ or $\Lambda$), 
slow modes emerge which prevent the system from
achieving its stationary state. 
As a consequence, during this neverending relaxation of the system, 
TT is broken
together with the possible additional symmetries which characterize
the stationary state.\footnote{The one described here is the typical
pattern of the spontaneous breaking of a symmetry. See, e.g.,~\cite{cu-rev}.}
This is what happens when ageing takes place in
critical systems with detailed balance. 
In the next subsection we explicitly show that this is also the case
for the contact process.

\subsection{The Gaussian approximation}
\label{sec-compu}

The analytic computation of the response function follows the
same steps as those discussed in~\cite{calabrese1} (see
also~\cite{oerding}) for the non-equilibrium relaxation of the Ising model
relaxing from a state with non-vanishing value of the magnetization. 
Here we briefly outline the calculation.
We assume that the initial state has a non-vanishing order parameter
$m_0 = \langle\p(x,t=0)\rangle$ which is spatially homogeneous, so
that the ensuing relaxation is characterized by translational
invariance in space. It is convenient to subtract from the order
parameter field $\p(x,t)$ its average value $m(t) =  \langle \p(x,t)
\rangle$, by introducing
\be
\psi(x,t) := \p(x,t) - m(t)\quad \mbox{and} \quad  \tilde{\psi}(x,t) := \tp(x,t)\,,
\label{defpsi}
\ee
leading to $\langle\psi(x,t)\rangle = 0$ during the relaxation.
The action $S$ [see equation~\reff{action}]
becomes, in terms
of these fields,
\begin{equation}
\label{action2}
S[\psi,\tilde{\psi}]  =  \int\rmd^dx\rmd t \left\{ \tilde{\psi}\left[\partial_t - \nabla^2 
+ 2 \sigma\right] \psi - \sigma \tilde{\psi}^2 - u 
(\tilde{\psi} - \psi) \tilde{\psi} \psi \right\} ,
\end{equation}
where $\sigma(t) \deff u\, m(t)$ satisfies the equation of motion
(no-tadpole condition)
\begin{equation}
\label{eqn_of_motion}
\partial_t \sigma + \sigma^2 + u^2 \langle \psi^2 \rangle = 0 .
\end{equation}
Equations~\reff{eqn_of_motion} and~\reff{action2} are the starting
point for our calculations. The Gaussian response and correlation
function are obtained by neglecting anharmonic terms in
equation~\reff{action2}:
\begin{eqnarray}
\label{resp_prop}
R_q^{\G}(t,s) &=& \theta(t-s) 
\exp\left\{-q^2(t-s) - 2 \int_s^t \rmd t' \sigma(t')\right\}\quad\mbox{and} \\
C_q^{\G}(t,s) &=& 2 \int_0^{t_<}\!\! \rmd t'\, R_q^{\G}(t,t') \sigma(t')
R_q^{\G}(s,t')
\end{eqnarray}
where $t_< \deff \min\{t,s\}$. [In what follows we denote the
results of the Gaussian approximation with the superscript $\G$.]
In particular,  solving~\reff{eqn_of_motion} with $u=0$  one finds
(see also~\cite{oerding} and the appendix of~\cite{ramasco})
$\sigma^{\G}(t) = (t+\sigma_0^{-1})^{-1}$, which scales according 
to~\reff{scalingm} ($\beta^{\G} =1$, $\nu
^{\G}=1/2$, $z^{\G} = 2$ and $\eta^{\G} =0$, see, e.g., \cite{zinn-justin}) 
with $A_m^{\G}= 1$, $B_m^{\G}=u$ and $\FF^{\G}_M(x) =
(1+x)^{-1}$. As explained in section~\ref{sec-scalingforms}, the leading
behavior for $t>s\gg \tau_m = (B_mm_0)^{-1/\varsigma} = \sigma_0^{-1}$ 
can be explored by taking the limit
$m_0\propto \sigma_0\rightarrow\infty$ from the very beginning. Accordingly,
$\sigma^{\G}(t) = t^{-1}$ and 
\begin{eqnarray}
R_q^{\G}(t,s) &=&\theta(t-s) \left( \frac{s}{t} \right)^2
\rme^{-q^2 (t-s)},\label{Rq0}\\  
C_q^{\G}(t,s) &=& 2\rme^{-q^2(t+s)} \left( t \,s\right)^{-2}
\int_0^{t_<}\!\! \rmd t'\, t'^3 \rme^{2 q^2 t'} , \label{Cq0}
\end{eqnarray}
which (for $q=0$) display the expected scaling
behaviors~\reff{scalRasy} and~\reff{scalCasy} with $\AAA^{\G}_R = 1$,
$\AAA_C^{\G}=1/2$ and $\FF^{\G}_{C,R}(x) \equiv 1$. 
Equations~\reff{fdrq} and~\reff{fdrqinf} yield $\FDRq^{\G}(t,s) =
\FDRq_\infty^{\G} = 2$, which is exact for $d>4$ but it is almost twice
as large as the result which was found in $d=1$, $\FDRq_\infty =
1.15(5)$~\cite{enss}. 
As explained in subsection~\ref{sec-scalingforms}, $\FDR^{-1} \neq 0$
is a signal of the breaking of the RR symmetry characterizing the
non-equilibrium stationary state of the CP (and of the DP universality
class in general). Here we show explicitly (within the Gaussian
approximation --- though the conclusion is expected to be valid also
beyond the approximation) 
that only the homogeneous fluctuation mode $q=0$ is, at
criticality, responsible for such a breaking, as in the case of
systems with detailed balance (see,
e.g,~\cite{cg-rev,calabrese3,calabrese1,cg-02a}). Indeed, let us
generalize equation~\reff{fdrq} to modes with $q\neq 0$:
\be
\FDRq^{-1}_q(t>s,s) \deff \frac{C_q(t,s)}{R_q(t,s)}\,.
\label{fdrqnz}
\ee
Within the Gaussian approximation one readily finds [see
equations~\reff{Rq0} and~\reff{Cq0}] 
\be
\left[\FDRq_q^{\G}(t,s)\right]^{-1}  = \left.
\frac{1}{2}\frac{4!}{y^4} \left( \rme^{-y}
-1 +  y - \frac{y^2}{2!} + \frac{y^3}{3!}\right)\right|_{y = 2 D q^2 s} ,
\ee
where the additional scaling variable $y = A_D D q^z s = D q^2 s$
appears. 
\begin{figure}[h]
\begin{center}
\includegraphics[scale=0.6]{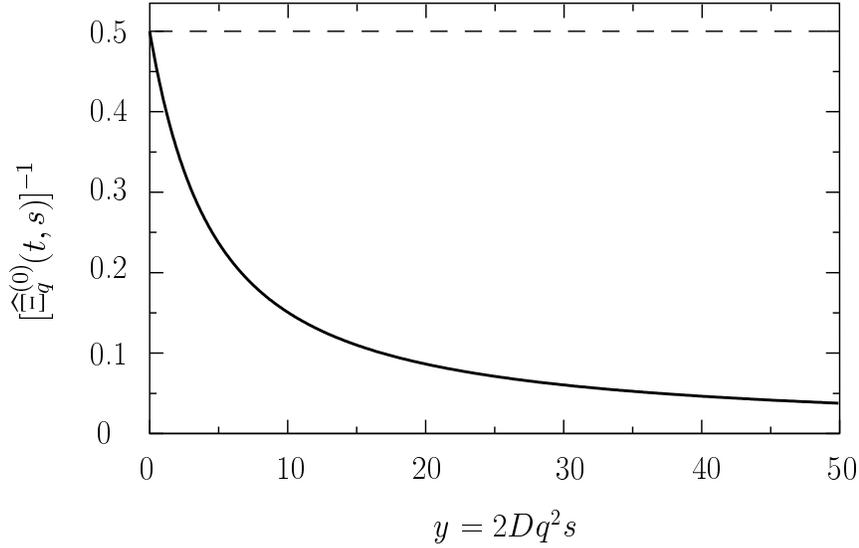}
\caption{\label{diss} Ratio $[\FDRq_q^{\G}(t,s)]^{-1}$ of
the correlation function to the response function [see
equation~\reff{fdrqnz}] within the Gaussian approximation (which
becomes exact for $d>4$). In the long-time limit $s\rightarrow\infty$,
$[\FDRq_q^{\G}(t,s)]^{-1}$ vanishes, unless $q=0$, in which
case it takes the value $1/2$.  $[\FDRq_q^{\G}]^{-1}
\neq 0$ signals the breaking of RR [see equation~\reff{eq-RR} and subsection~\ref{sec-relX}].} 
\end{center}
\end{figure}
%
%
%
This expression --- as the Gaussian fluctuation-dissipation ratio for
systems with detailed balance (see, e.g, \cite{cg-02a}) --- 
is independent of $t$
and depends on $y$ only. 
In particular,
in the long-time limit $s
\gg q^{-2}$ 
one has $\left[\FDRq_q^{\G}(y \gg
1)\right]^{-1} \sim  1/y \rightarrow 0$, for any mode with $q\neq 0$, 
indicating that the RR
symmetry is asymptotically realized. On the other hand, for $q=0$, 
$\left[\FDRq_q^{\G}(y = 0)\right]^{-1} =1/2$, independently of $s$. 
For a quench into the active phase $\Lambda > \Lambda_c$ the action
$S$ and the propagators~\reff{Rq0} and~\reff{Cq0} get modified
according to  $q^2 \rightarrow q^2 +r$, with $r>0$  (see, e.g.,
\cite{tauber,janssen}). Therefore in this case, 
$y = 2 D (q^2+r) s$, yielding 
$\left[\FDRq_q^{\G}(y \gg
1)\right]^{-1} \rightarrow 0$ for $s \gg r^{-1}$ and independently of
$q$, in agreement 
with what has been stated in subsection~\ref{sec-relX} and with the
available numerical evidences~\cite{enss}.

\section{The response function}
\label{sec-RF}

In this subsection we determine the response function up to first
order in $\eps = 4-d$ ($4$ being the upper critical
dimension $d_{\rm ucd}$ of the model~\cite{tauber}, 
above which the Gaussian results
become exact). 
Some of the details of the calculation are provided in the appendix. 

For future reference we recall that the critical exponents of the DP
universality class are~\cite{janssen,hinrichsen1,tauber} 
$\eta = -\eps/6 + O(\eps^2)$, $z = 2 - \eps/12 +
O(\eps^2)$, $\beta/\nu = (d+\eta)/2$ and therefore [see equation~\reff{expbt}]
\be
\ha = -\frac{\eta}{z} = \frac{\eps}{12} + O(\eps^2)
\quad\mbox{and}\quad \bt = -2 + \frac{\eps}{6} + O(\eps^2).
\label{exp1l}
\ee
The expression for the renormalized response function is (where
$x\deff s/t \le 1$)
\begin{eqnarray}
\label{response_final}
\hspace{-1cm}
R_{R,q=0} (t,s) &=& x^2\left\{ 1 + \eps \left[
-\frac{1}{4} \ln x + \frac{1}{12} \ln s \nonumber
\right.\right. \\
&+& \left.\left.
\frac{\pi^2}{12} + \frac{\ln(1-x)}{x} - \frac{1}{2}
\mbox{Li}_2(x) - \frac{11}{12} \ln(1-x) + \frac{x}{12} 
\right] \right\} + O(\eps^2),
\end{eqnarray}
(see the appendix for details)
which can be cast in the expected scaling form~\reff{scalRasy} with
\begin{eqnarray}
&&\AAA_R = 1 - \eps \left( 1 - \frac{\pi^2}{12}\right) + O(\eps^2) ,
\label{AR}\\
&&\FF_R(x) = 1 + \eps \left[ 1 + \frac{x}{12}  + \left(\frac{1}{x} -
1\right) \ln(1-x) - \frac{1}{2} \mbox{Li}_2(x)  \right] + O(\eps^2),
\label{FR}
\end{eqnarray}
[note that $\FF_R(0) = 1$ and 
$\FF_R(x\rightarrow 1^-) = 1 + \eps (13 - \pi^2)/12 + O(\eps^2)$]
and the proper exponents~\reff{exp1l}.

\subsection{Comparison with LSI}
\label{sec-LSI}

For the response function, LSI provides the following prediction
\cite{henkel06}: 
\be
\label{lsi_prediction}
R_{q=0}(t>s,s) = r_0 s^A x^B (1-x)^C ,
\label{RLSI}
\ee 
where $x\deff s/t$ and $A$, $B$, $C$ and $r_0$ are free parameters. In a
first version of LSI it was assumed  $A=C$ but more recently
it has been suggested that this constraint can be
relaxed~\cite{picone,henkel06}. 
On the other hand, requiring $R_{q=0}$ to have the
correct scaling dimension (see
subsection~\ref{sec-scalingforms}), leads to $A = \ha =
-2\delta+d/z$. 
In addition, the
well-established short-time behavior
$s/t \rightarrow 0$, from equation~\reff{scalRasy}, leads to $B =
-\bt - \ha = 1 + \delta$.
The last free parameter, $C$, can be in principle fixed
by requiring the response function to display 
a {\it quasi-stationary} regime (analogous to the quasi-equilibrium
regime in systems with detailed balance, see, e.g., Section 2.5
in~\cite{cg-rev}) for $\Delta t \ll s$ where $\Delta t \deff t-s$,
in which the behavior of the system depends only on $\Delta t$ and
not on $s$. 
This would require $C=A$~\cite{hinrichsen2}.
In turn,
comparing with equation~\reff{scalRasy}, this would imply $\FF_R(x) = 1$,
$\forall x \in [0,1]$ which is in contrast with the analytical
expression in equation~\reff{FR}. 
The discrepancy between the actual
scaling behavior of the CP and the prediction of LSI with $A=C$,
was originally overlooked in~\cite{enss} but then it became apparent
for  $t/s \lesssim 2$ in the detailed numerical analysis carried out
in~\cite{hinrichsen2}, which is
also supported by our analytical result. 
On the other hand, if one questions the existence of such a
quasi-stationary regime in
the CP, $C$ remains an available fitting parameter. By adjusting it
the agreement between
numerical data and LSI improves significantly~\cite{henkel06}, down to
$t/s - 1 \simeq 10^{-1}$, being discrepancies observed only for smaller  $t/s
- 1$~\cite{hinrichsen2}. 
It has been suggested~\cite{henkel06} 
that such discrepancies might be caused by 
corrections to scaling due to having $t-s$ of the order of microscopic
time scales when exploring the regime $t/s-1 \simeq 0$ with finite $s$ 
 ---  However, this does not seem to be the case
for the data reported in~\cite{hinrichsen2}.

\begin{figure}[h]
\begin{center}
\includegraphics[scale=0.85]{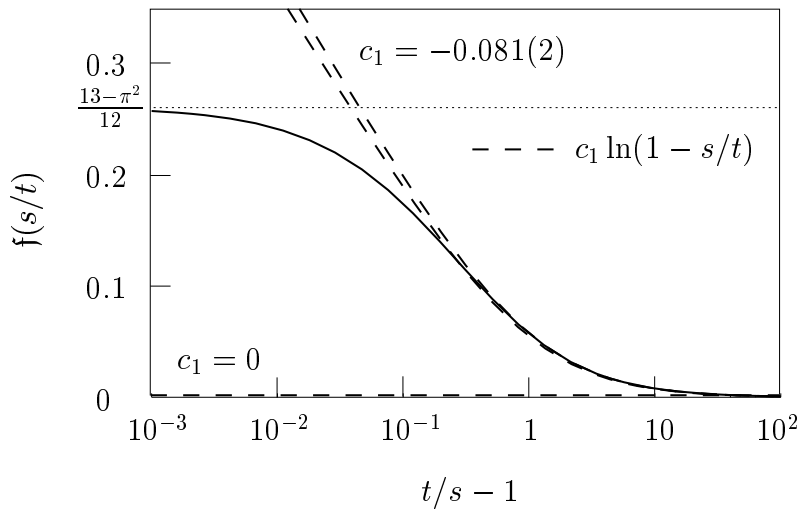}
\caption{\label{DPLSI}Comparison between $\eps^{-1}\ln \RR_\ft(x)$
and $\eps^{-1}\ln \RR_\lsi(x)$ for two different values of the
fit-parameter $c_1$.}
\end{center}
\end{figure}

In order to compare the analytic expression of $R_{q =0}(t,s)$ to the
prediction of LSI with $C\neq A$ we introduce the quantity 
$\RR(x) \deff \AAA_R^{-1}(t-s)^{-\ha}(t/s)^{-\bt} R_{q=0}(t,s)$
where $\AAA_R$ is fixed by the requirement $\RR(x\rightarrow 0) = 1$. 
Accordingly, the field-theoretical
prediction~\reff{scalRasy}, yields (for $d<4$)
\be
\RR_\ft(x) = \FF_R(x) = 1 + \eps\,
\f(x) + O(\eps^2)
\label{RR-FT}
\ee 
where $\f(x)$ is given by the expression in square brackets in
equation~\reff{FR}:
\be
\f(x) \deff  1 + \frac{x}{12}  + \left(\frac{1}{x} -
1\right) \ln(1-x) - \frac{1}{2} \mbox{Li}_2(x) ,
\label{ffr}
\ee
with $\f(1) = (13-\pi^2)/12$. LSI predicts, instead,
\be
\RR_\lsi(x) = (1 - x)^{\Delta C}
\label{RR-LSI}
\ee
where $\Delta C \deff C - A$. 
Above the upper critical dimension $d_{\rm ucd} = 4$,
$\RR_\ft(x) \equiv 1$ and $\RR_\lsi$ 
fit it with  $\Delta C = 0$. Assuming continuity of
critical exponents, one expects for $d<4$ (i.e., $\eps >0$), $\Delta
C = c_1 \eps + O(\eps^2)$, --- where $c_1$ is a fitting coefficient
--- and therefore $\RR_\lsi(x) = 1 + \eps \, c_1
\ln (1-x) + O(\eps^2)$. Of course $\RR_\ft - \RR_\lsi = \eps[\f(x) -
c_1 \ln(1-x)] + O(\eps) \neq 0$ for $\eps > 0$ (i.e., $d<4$) 
whatever
the choice of $c_1$ is, as it is clear by inspecting the
expression~\reff{ffr}. Strictly speaking LSI in its present form 
{\it is not} a symmetry
of $R_{q=0}(t,s)$, whatever $\Delta C$ is, at least sufficiently close
to $d_{\rm ucd}$ (small $\eps$). 
Nonetheless, the shape of
$\f(x)$ is such that a proper choice of $c_1$ can reduce the
difference $\RR_\ft - \RR_\lsi$.
In figure~\ref{DPLSI} we report the comparison between $\ln \RR_\ft(x) =
\eps \f(x) + O(\eps^2)$ and $\ln \RR_\lsi(x) = \eps\, c_1\ln(1-x) +
O(\eps^2)$ with $c_1 = 0$ (i.e., $C = A$) and $c_1 = -0.081(2)$
(corresponding to $ C \neq A$). 
%
This plot clearly indicates that the scaling function $\RR_\ft(x)$ shows
corrections to the behavior predicted by LSI with $C=A$
already for $t/s-1 \lesssim 8$, i.e., $0.11\lesssim x \le 1$, for which
$\RR_\ft(x)/\RR_\lsi(x) \gtrsim 1 + 0.01 \eps + O(\eps^2)$. On the
other hand the same discrepancy is observed only for $t/s-1 \lesssim
0.2[0.13]$, i.e., $0.83[0.88]\lesssim x \le 1$ for $c_1 = -0.083
[-0.079]$ and therefore a suitable choice of $c_1$ (i.e., $C$)
reduces by more than one order of magnitude the value of $t/s-1$ below
which the difference
between the predicted scaling
behavior and LSI becomes sizable.
This is analogous to what has been observed 
in numerical data~\cite{hinrichsen2,enssprivate} 
for the scaling of the autoresponse
function $R_{x=0}(t,s)$. (See, e.g., figure 2
in~\cite{hinrichsen2}.) 
On the other hand, the field-theoretical results presented here are
free from numerical artefacts and corrections to scaling, which were
invoked in~\cite{henkel06} to explain the findings
of~\cite{hinrichsen2}.

As a result of our fitting procedure, we have determined the
``$\eps$-expansion'' of the exponent $\Delta C$, finding $\Delta C =
-0.081(2) \eps + O(\eps^2)$. According to the notations adopted 
in~\cite{henkel06} $\Delta C = - (a'-a)$ 
and therefore we conclude that $a'-a = + 0.081(2) \eps +
O(\eps^2)$. For the one-dimensional contact process it was found 
$a'-a = + 0.270(10)$~\cite{henkel06,hinrichsen2}\footnote{Note that
in~\cite{hinrichsen2} the value of $a'$ is reported with the wrong
sign.}, whose sign is indeed in agreement with our result (generally
speaking, the sign of the correction to the Gaussian behavior is
correctly captured by the lowest non-trivial order in the standard
$\eps$-expansion). Insisting in interpreting our result as an
$\eps$-expansion we can even attempt an estimate of $a'-a$ in physical
dimensions $\eps = 1$, $2$ and $3$. It comes as a surprise that not
only the order of magnitude is the correct one but also the estimate for the
one-dimensional CP (i.e., $\eps=3$) $a'-a \simeq +0.24$ is in very
good agreement with the actual numerical result. 

In passing, we note that a mechanism similar to what has been
described here 
could possibly explain the (apparent) agreement
between LSI with $C\neq A$ and the scaling form of the integrated
response function of the magnetization in the Ising model quenched
from high temperatures to the critical
point~\cite{pleimling,henkel06}.

\subsection{One-loop prediction for $\FDR$}

The correlation function was already calculated in~\cite{oerding}, 
up to one loop in the $\eps$-expansion, with the result
\be
2 \, \AAA_C = A \times \left[ 1 - \frac{\eps}{6} F(\infty)\right] + O(\eps^2) = 1 + \frac{\eps}{6} \left( \frac{9 \pi^2}{20} - \frac{361}{80}\right) + O(\eps^2)
\label{AC}
\ee
and
\be
\FF_C(x) = 1 - \frac{\eps}{6}\left[F(x^{-1}) - F(\infty)\right] + O(\eps^2)
\label{FC}
\ee
where $A$ and $F$ are given, respectively, in equation~(4.4) and (3.5)
therein. 
Accordingly,
we obtain for the ratio $\FDRq_\infty$ in~\reff{fdrqinf} the value
\begin{equation}
\label{fdr_result}
\FDRq_\infty =2 \left[ 1 - \epsilon\left( \frac{119}{480} - \frac{\pi^2}{120}\right) \right] + O(\eps^2)\;.
\end{equation}
Clearly the Gaussian value of $\FDRq_\infty$ gets a downward
correction ($119/480 - \pi^2/120 \simeq 0.166$) for $\eps
> 0$. 
This is qualitatively satisfying, as it brings 
the theoretical estimate closer to the value which was
determined numerically for $d=1$ [$\FDR_\infty =
1.15(5)$]. In addition, a direct extrapolation of
equation~\reff{fdr_result} to $\eps=3$ (neglecting possible
$O(\eps^2)$ terms) yields
$\FDRq_\infty^{d=1} = 1.01$ which is --- for a one-loop calculation
--- in a surprisingly good agreement with the actual
value.\footnote{Extrapolating the result~\reff{fdr_result} as
$\FDRq_\infty^{\rm (extr)} = 2/[1 + \eps (119/480 - \pi^2/120)]$ would
give $1.33$ for $\eps=3$, still not too far from the numerical result.}
More reliable theoretical estimates can only be based on the knowledge
of higher-loop contributions.

\section{Conclusions}
\label{sec-concl}
In this note we have studied the ageing behavior in the contact
process by adopting a field-theoretical approach. The results
presented here equally apply to all the models belonging to the same
universality class as the contact process, such as the directed 
percolation.

We confirmed analytically the scaling behavior 
\begin{equation}
R_{x=0}(t,s) = s^{-1-a} f_R(t/s), \qquad C_{x =
0}(t,s)= s^{-b} f_C(t/s) ,
\end{equation}
which was observed numerically for the two-time response and
correlation functions~\cite{enss,ramasco} after a quench from the
active phase to the critical point. 
In addition we have shown that:
\vspace{-0.7cm}
\begin{itemize}
\item[(a)] The relation $b=1+a$, due to the rapidity-reversal
symmetry, and $\lambda_R = \lambda_C$ [see
equation~\reff{scaling_forms2}],  
hold beyond the numerical evidences provided in~\cite{enss,ramasco}.
\item[(b)] $\lambda_{R,C}$ are related to known static
and dynamic exponents via $\lambda_{R,C} = z (1 +\delta) + d$
[see equation~\reff{eq-exp}], which is
confirmed by the available numerical data 
(see table~\ref{tabexp}).
\item[(c)] The ratio $\FDR$ of the response to the
correlation function [equation~\reff{fdr}] defines a {\it universal
function} and its long-time limit $\FDR_\infty$ a {\it universal
amplitude ratio}, which can therefore be studied within 
a field-theoretical approach. 

The result of the $\eps$-expansion
around $d=4$ [see equation~\reff{fdr_result}] 
is in qualitative agreement with the numerical estimate
of $\FDR_\infty^{d=1}$~\cite{enss}. 
\item[(d)] After a quench, $\FDR_\infty^{-1} \neq 0$ for the 
CP, like $X_\infty -
1 \neq 0$ for systems satisfying detailed balance, is a signature of 
the spontaneous breaking of the symmetry $\mathfrak S$ associated with
the corresponding non-equilibrium ($\mathfrak S$  = rapidity-reversal)
and equilibrium ($\mathfrak S$  = time-reversal) stationary states.
\item[(e)] The {\it universal} scaling function $\FF_R(x)$ --- describing
the ageing behavior of the response function $R_{q=0}(t,s)$ [see
equation~\reff{scalRasy}] --- {\it does not} agree with the prediction
of the theory of local scale invariance when the non-Gaussian
fluctuations are taken into account [specifically already at $O(\eps)$
for $\eps > 0$, see equation~\reff{FR}]. 
This provides an additional example against the general applicability
of the present form of LSI to critical systems. 
It remains to be seen whether an extension of LSI to systems with
nonvanishing order parameter will be able to describe correctly
the scaling function of the CP.
\end{itemize}

{\bf Acknowledgments:} 
We thank Tilman Enss, Malte Henkel, Michel Pleimling, Beate
Schmittmann and Uwe T\"auber for useful discussions and comments and
the organizers of the school {\it Non-Equilibrium Dynamics of
Interacting Particle Systems} at the Isaac Newton Institute of Mathematical
Sciences, Cambridge, where this work was initiated.
FB acknowledges the support by the 
Deutsche Forschungsgemeinschaft  through grant no. PL 323/2.

\appendix

\section{Diagrammatic expansion of the response function}
\begin{figure}[t]
\centering\includegraphics[scale=1]{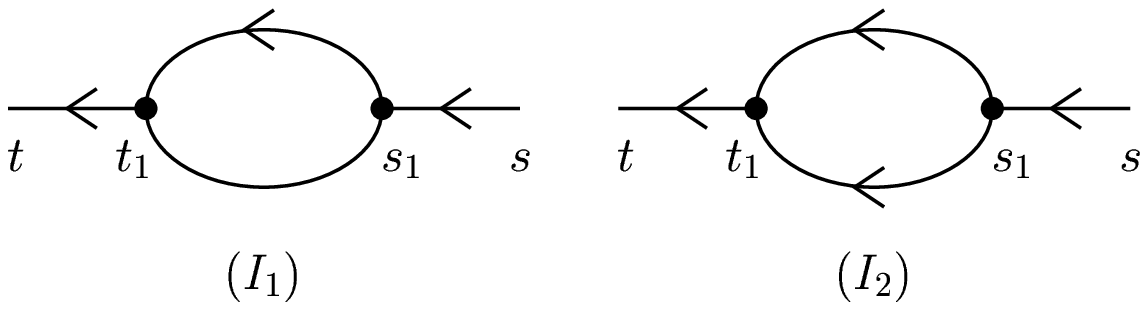}
\caption{\label{figura}Diagrammatic contributions to the
response function up to first order in $v$. Directed and undirected 
lines represent response~\reff{Rq0} and correlation~\reff{Cq0}
propagators, respectively.}
\end{figure}

In this appendix we provide some details of the calculation of the
response function. First of all we solve equation~\reff{eqn_of_motion}
 --- to first order in the effective coupling constant $\tilde{v} :=
(8 \pi)^{-d/2}u^2/D^2$ --- where $\langle \psi^2 \rangle$ is given by
\be
\langle \psi^2(x,t)\rangle = \int\!\!(\rmd q) C^{(0)}_q(t,t) + O
(\tilde{v}) = (8 \pi)^{-\frac{d}{2}}
 r_d t^{-\frac{d}{2}} + O(\tilde{v})\;.
\ee
For later convenience we introduce the notation $(\rmd q) \deff \rmd^d
q/(2\pi)^d$ [and dimensional regularisation of the integrals is
understood] and $r_d = 12\Gamma(1-d/2)/\Gamma(5-d/2) =
-12/\eps + 3 + O(\epsilon)$. Accordingly, $\sigma(t)$ is given by
\be
\label{result_sigma}
\sigma(t) = \sigma^{(0)}(t) \left (1-\tilde{v} \frac{r_d}{\epsilon/2
+1} t^{\epsilon/2}\right) + O(\tilde{v}^2) \;,
\ee
where $\sigma^{(0)}(t) = 1/t$ is the solution within the Gaussian
approximation. (As we are interested in the
long-time limit, we have taken $\sigma_0 \rightarrow \infty$.)
As expected, the expression~\reff{result_sigma} has a dimensional pole
for $\eps\to 0$, which is removed once the renormalized quantities
(with subscript $R$) are introduced. 
In particular (following~\cite{tauber}):
\be
\psi_R = Z_\psi^{1/2} \psi, \quad \tilde{\psi}_R =
Z_\psi^{1/2}\tilde{\psi}, \quad D_R = Z_D D, \quad\mbox{and}\quad \tilde{v}_R =
Z_{\tilde{v}} \mu^{-\eps} \tilde{v},
\label{ren}
\ee
where $\mu$ is an arbitrary momentum scale and
$Z_\psi = 1 - 4\tilde{v}/\eps + O(\tilde{v}^2)$, 
$Z_D =  1 + 2\tilde{v}/\eps + O(\tilde{v}^2)$ and
$Z_{\tilde{v}} =  1 - 24\tilde{v}/\eps + O(\tilde{v}^2)$ are suitable
renormalization constants. Taking into
account the definition of $\sigma$: 
$\sigma(t) = (8\pi)^{d/4} D \tilde{v}^{1/2} m(t)$ and that $m_R
= Z_\psi^{1/2} m$, one finds for the renormalized function
$\sigma_R(t)$ at the fixed-point $\tilde{v}^*_R = \eps/24 +
O(\eps^2)$~\cite{tauber}
\be
\sigma_R(t) = \frac{1}{t}\left(1+\frac{\eps}{4} \ln t - 9 \eps
\right) + O(\eps^2)\,,
\ee
which scales  according the long-time limit of
equation~\reff{scalingm} with the proper exponents [see before
equation~\reff{exp1l} in section~\ref{sec-RF}] 
and $A_m/B_m = [1 - 9\eps + O(\eps^2)]/u^*$.
To compute the response function to first order in
$\tilde{v}$, one has to take into account that the Gaussian propagator
gets modified because of the $O(\tilde{v})$-term in
equation~\reff{result_sigma}~\cite{calabrese1}.
Using~\reff{result_sigma} 
in equation~\reff{resp_prop}, one
gets (hereafter $t>s$ and $x\deff s/t$)
\be
R'^{(0)}_{q = 0}(t,s) = x^2 \left[ 1 + \tilde{v}
s^{\epsilon/2}\left( \frac{24}{\epsilon} \ln x - 18
\ln x - 6 \ln^2 x \right) + O(\tilde{v}^2,\eps\tilde{v}) \right]\;.
\ee
In addition, there are two diagrams contributing to the
response function at order $\tilde{v}$, 
depicted in figure~\ref{figura}. Notice that we can use the
$O(\tilde{v}^0)$-propagators to compute them up to $O(\tilde{v})$.
We first need the expressions for the following bubbles
\begin{eqnarray}
B_{RC}(t_1,s_1) &\deff& \int\!\!(\rmd q)\, R_{q}^{(0)}(t_1,s_1)
C_{q}^{(0)}(t_1,s_1) \nonumber \\
& & = 2 (8 \pi)^{-\frac{d}{2}} t_1^{-4} \int_0^{s_1} \rmd t'\, t'^3
(t_1-t')^{-\frac{d}{2}} \\
B_{RR}(t_1,s_1) &\deff& \int\!\!(\rmd q)\,
\left[R_{q}^{(0)}(t_1,s_1)\right]^2
= (8 \pi)^{-\frac{d}{2}}\left( \frac{s_1}{t_1} \right)^4
(t_1-s_1)^{-\frac{d}{2}}
\end{eqnarray}
for $t_1 \leq s_1$, whereas for $s_1 > t_1$ both $B_{RC}$ and $B_{RR}$
vanish because of causality. The contribution $I_1$ is then
given by the integral
\be
I_1 = \int_s^t \rmd t_1\, \int_s^{t_1}\rmd s_1\, R_{q=0}^{(0)}(t,t_1)
B_{RC}(t_1,s_1) R_{q=0}^{(0)}(s_1,s) 
\label{II1}
\ee
which has essentially been calculated in~\cite{calabrese1}. (Indeed,
taking into account that $B_{RC}(t_1,s_1)$ here is equal to $t_1^{-1}
H_d(t_1,s_1)$ of~\cite{calabrese1} --- see equation~(A.4) therein ---
one finds that the rhs of equation~\reff{II1} is equal to
$(s/t)^{1/2}I_2(t,s)/3$ where $I_2$ is given by equation~(A.11)
of~\cite{calabrese1}.)
The result is 
\begin{eqnarray}
\hspace{-2.0cm}
I_1 = (8 \pi)^{-\frac{d}{2}} x^2 s^{\eps/2}& \Big[
-\frac{4}{\epsilon} \ln x - 1+x +
\frac{2 \pi^2}{3} + 6 \ln(1-x)\left(\frac{1}{x}-1\right) +
3 \ln x \nonumber \\ 
& +   \ln^2 x - 4 \mbox{Li}_2(x) + O(\epsilon) \Big] \;.
\end{eqnarray}
The contribution $I_2$ is instead given by
\begin{eqnarray}
I_2 &\deff& \int_s^t \rmd t_1 \int_s^{t_1}\rmd s_1 R_{q=0}^{(0)}(t,t_1)
B_{RR}(t_1,s_1) R_{q=0}^{(0)}(s_1,s) \nonumber \\ &=&
(8 \pi)^{-\frac{d}{2}} x^2 \int_s^t \rmd t_1\, t_1^{-2} \int_s^{t_1}
\rmd s_1\,
s_1^2 (t_1 -s_1)^{\epsilon/2-2}\;.
\end{eqnarray}
We compute first the integral over $s_1$, which yields after some
manipulations and for generic $\epsilon$:
\begin{eqnarray}
\int_s^{t_1}\!\!\rmd s_1 \, s_1^2 (t_1-s_1)^{\epsilon/2-2} &= 
\frac{(t_1-s)^{\epsilon/2-1} s^2}
{\epsilon/2+1} + \frac{2
(t_1 -s)^{\epsilon/2-1}t_1 s}{(\epsilon/2-1)
(\epsilon/2+1)}
\\ &+ \frac{2(t_1 -s)^{\epsilon/2} t_1 }{(\epsilon/2-1)
\epsilon/2 (\epsilon/2+1)} \nonumber\;.
\end{eqnarray}
The resulting three integrals over $t_1$ are easy and one finally gets:
\begin{eqnarray}
I_2 & = & (8 \pi)^{-\frac{d}{2}} x^2 s^{\eps/2} \Big[
-\frac{2}{\epsilon} + \frac{4 \ln x}{\epsilon} - \ln^2 x
+ \frac{\pi^2}{3} \nonumber  \\ & + & 2 \mbox{Li}_2(x) + x -
\ln(1-x)  - 2 \Big]
\end{eqnarray}
where the relation $
\mbox{Li}_2(1-x^{-1}) =
-\ln^2 x/2 - \pi^2/6 + \ln x \ln(1-x)
+ \mbox{Li}_2(x)
$ was used. Collecting the different contributions we get the final
expression for the response function:
\begin{eqnarray}
R_{q=0}(t,s)  & = & R_{q=0}'^{(0)}(t,s) + 4 vI_1 - 2
v I_2  + O(v^2,\eps v)\\  & = &  
x^2\left\{ 1 + \tilde{v} s^{\eps/2}\left[\frac{4}{\epsilon}- 6 \ln x + 2\pi^2
- 12 \mbox{Li}_2(x) \right.\right. \nonumber \\ & & \left.\left.-
22\ln(1-x) + 2 x + \frac{24 \ln(1-x)}{x}
\right]\right\} + O(\tilde{v}^2,\eps\tilde{v}) \,.
\end{eqnarray}
Its dimensional pole is properly removed by
introducing the renormalized quantities according to
equation~\reff{ren}, so that $R_{R,q=0} = Z_\psi R_{q=0}$ which 
gives indeed the expression in equation~\reff{response_final} for
$\tilde{v} = \tilde{v}_R^* = \eps/24 + O(\eps^2)$.


\end{document}